# A new mechanical stellar wind feedback model for the Rosette Nebula

C. J. Wareing,[1][*] J. M. Pittard,[1] N. J. Wright[2] and S. A. E. G. Falle[3]

[1]*School of Physics and Astronomy, University of Leeds, Leeds LS2 9JT, UK*
[2]*Astrophysics Group, Keele University, Keele ST5 5BG, UK*
[3]*School of Mathematics, University of Leeds, Leeds LS2 9JT, UK*

**ABSTRACT**

The famous Rosette Nebula has an evacuated central cavity formed from the stellar winds ejected from the 2–6 Myr old codistant and comoving central star cluster NGC 2244. However, with upper age estimates of less than 110 000 yr, the central cavity is too young compared to NGC 2244 and existing models do not reproduce its properties. A new proper motion study herein using *Gaia* data reveals the ejection of the most massive star in the Rosette, HD 46223, from NGC 2244 occurred 1.73 (+0.34, −0.25) Myr (1$\sigma$ uncertainty) in the past. Assuming this ejection was at the birth of the most massive stars in NGC 2244, including the dominant centrally positioned HD 46150, the age is set for the famous ionized region at more than 10 times that derived for the cavity. Here, we are able to reproduce the structure of the Rosette Nebula, through simulation of mechanical stellar feedback from a 40 $M_\odot$ star in a thin sheet-like molecular cloud. We form the 135 000 $M_\odot$ cloud from thermally unstable diffuse interstellar medium (ISM) under the influence of a realistic background magnetic field with thermal/magnetic pressure equilibrium. Properties derived from a snapshot of the simulation at 1.5 Myr, including cavity size, stellar age, magnetic field, and resulting inclination to the line of sight, match those derived from observations. An elegant explanation is thus provided for the stark contrast in age estimates based on realistic diffuse ISM properties, molecular cloud formation and stellar wind feedback.

## 1 INTRODUCTION

The stellar winds and ionizing radiation forming the Rosette Nebula, a region of ionized gas in a giant molecular cloud in the constellation of Monoceros, come from the equidistant star cluster NGC 2244, located in the central cavity of the Nebula. NGC 2244 has approximately 2000 members (Wang et al. 2008), including five massive (O-type) stars. The most massive are HD 46150 and HD 46223. Both of these stars have winds that are two orders of magnitude more powerful than even the other O-type stars in the Nebula (Howarth & Prinja 1989). HD 46150 is positioned roughly at the centre of the cavity and is close to the centre of the cluster (Wang et al. 2008). However, HD 46223 appears to lie at the edge of the cavity, and has few cluster companions. In spite of this, proper motion analysis in the literature to date does not suggest ejection of HD 46223 from the cluster centre (Zacharias et al. 2004). The central position and relative wind strength of HD 46150 suggest that it alone has dominated the formation and evolution of the Nebula.

The cluster stars are 2–6 Myr old (Bonatto & Bica 2009), although all five massive stars appear to have ages of around 2 Myr (Martins et al. 2012). It should be noted that it is common for studies to find ages of around 2 Myr for groups of massive stars when they are in fact older, either because the studies use non-rotating massive star models (as opposed to more accurate rotating models, e.g. Ekström et al. 2012) or due to the bias that older massive stars have died. A discussion of this effect can be found in Wright, Drew & Mohr-Smith (2015), concerning the massive stars in Cyg OB2. HD 46150 and HD 46223 are thought to be 50–60 $M_\odot$ (Martins et al. 2012), although some estimates are lower, at around 35 $M_\odot$ for HD 46150 (Mahy et al. 2009). Current theoretical understanding of massive star evolution implies that at 2 Myr, they are around half-way through their evolution on the main sequence, and that their winds have experienced only minor variation over this time (Ekström et al. 2012). Evolutionary thinking therefore implies that the Rosette Nebula has taken approximately 2 Myr to form. We show a recent observation of the Rosette Nebula taken as part of the INT Photometric H$\alpha$ Survey (IPHAS) of the Northern Galactic Plane (Drew et al. 2005) in Fig. 1(a).

Recent work has highlighted a discrepancy between the age of the stars in NGC 2244 and the estimated age of the Nebula based on the measured expansion velocities and theoretical models of the wind-blown bubble dynamics (Bruhweiler et al. 2010). Therein, the authors found that in the case of the Rosette Nebula, the evacuated central cavity has a much smaller size than modelling (Freyer, Hensler & Yorke 2003, 2006) would predict. They found an age of 64 000 yr for the bubble – a remarkable discrepancy compared to the stellar ages. Varying dynamical models and a number of parameters, including reducing the central luminosity by a factor of 20, yielded an upper age limit of the bubble of 170 000 yr. Using the Weaver et al. (1977) analytical model for adiabatic bubbles, they estimated an age of 76 000 yr and a shell expansion velocity of 48 km s$^{-1}$. A momentum-conserving model instead yielded an age estimate of 270 000 yr and a shell velocity of around 12 km s$^{-1}$. Given the observed shell expansion velocity of 56 km s$^{-1}$ and upper dynamical limit of 110 000 yr, Bruhweiler et al. (2010) find the adiabatic case more consistent with the observations. They went on to postulate that the H II region surrounding the central cavity was a result of a recent ejection event in the history of the most massive stars in NGC 2244, but this is difficult to understand in terms of the evolutionary stage of HD 46150 and HD 46223. Bruhweiler et al. also 'emphasize that in the case of the Rosette we cannot rule out that there is an asymmetric cavity where the much larger axis of the cavity is directed towards the observer'. They note this seems like a remote possibility, but that it 'could also explain the small radius that we see in the plane of the sky'. Such a scenario would require some method of focusing the stellar wind in order to form an asymmetric structure, with a ratio of major to minor axes greater than 17, which Bruhweiler et al. found 'uncomfortably large' and 'extremely unlikely'. Rejecting the age estimate of 64 000 yr and adopting an H II region expansion velocity of 13 km s$^{-1}$ from the observational data presented by Celnik (1985), leads to a simple dynamical upper age limit of 450 000 yr, at least four times less than the stellar age estimates of the members of NGC 2244. In fact, Celnik was also forced to adopt a complex morphology after finding single and multiple shell models did not fit the radio observations; a single shell model with a hole towards the observer was found to provide the best fit, but further improvements required the assumption of spherical symmetry to be dropped. Celnik did not find a clear origin of the hole conveniently aligned with the line of sight.

[*] E-mail: C.J.Wareing@leeds.ac.uk

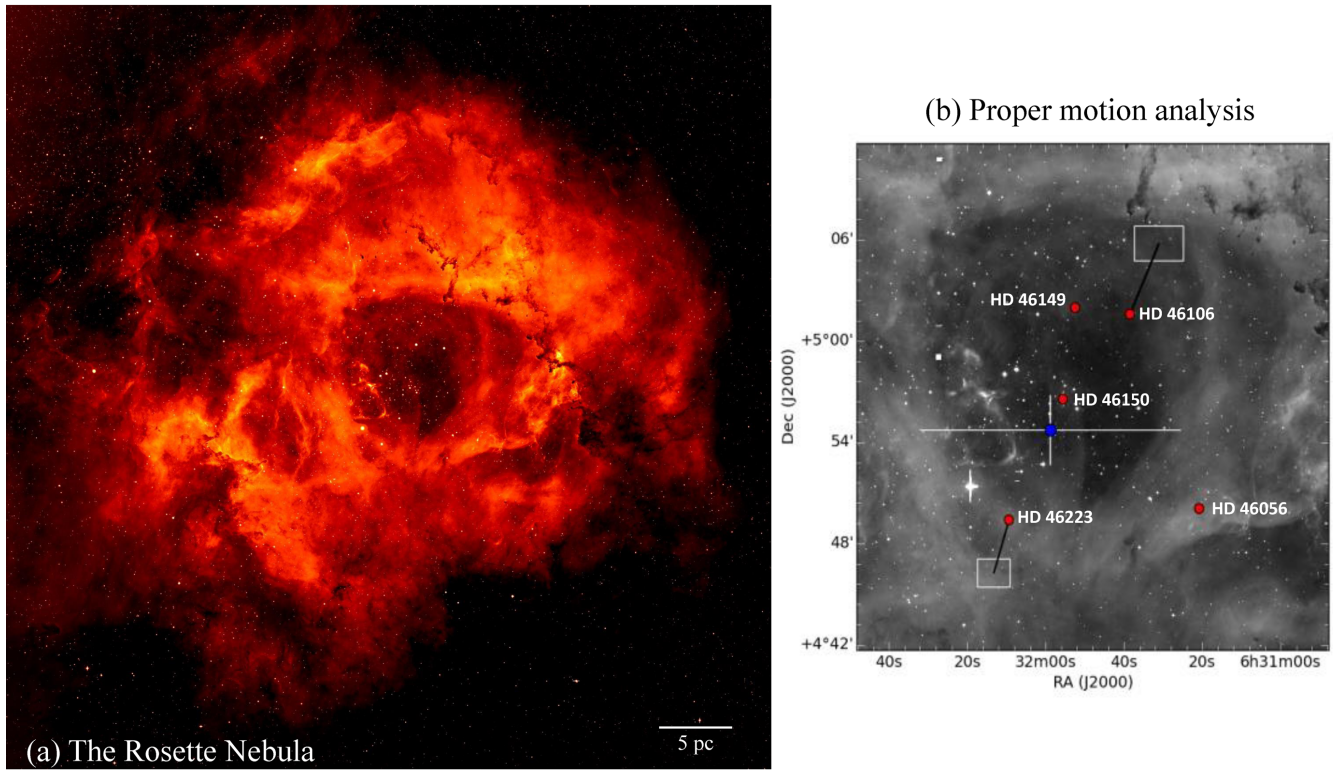

**Figure 1.** (a) Three-colour image (RGB = Hα, *r*, *i*) of the Rosette Nebula prepared from data from the IPHAS (Drew et al. 2005). (b) Proper motion analysis. *Hipparcos* and Tycho members in the *Gaia* DR1 catalogue are shown as red points. The proper motion vectors of the two runaway stars, after subtracting the weighted-mean motion of the cluster stars, are shown with black lines. Vector lengths represent the motion in 1 Myr, with proper motion uncertainties shown as white boxes around the endpoint of the vectors. The best-fitting back-traced 'interaction' point for these two stars is shown as a blue circle, with 1σ error bars in white. The background image is the main figure. Proper motion data are available from https://doi.org/10.5518/311.

Schneps, Ho & Barrett (1980) observed globules blueshifted by ∼17 km s$^{-1}$ with respect to the mean rest velocity of the H II region, coinciding with the well-known bulk radial velocity (of ∼20 km s$^{-1}$) observed in the Rosette. The globules are being swept away from NGC 2244 together with ionized gas, consistent with the entire structure being driven by stellar winds. The elephant-trunk globules are a result of stellar photoionization and wind action, and they deduced a formation time-scale on the order of 2.6–5.4 × 10$^5$ yr, much closer to the upper estimates of Bruhweiler et al. than the 2 Myr age of the stars. Dent et al. (2009) studied a much larger region of the Rosette and a much larger number of compact clumps, finding time-scales on the order of 10$^6$ yr and evidence that many of the clumps are being accelerated towards us. However, the near-constant velocity gradients that they found are difficult to explain in the context of radiatively driven clump acceleration. In addition, they find evidence that many of the clumps lie in an inclined molecular ring with a dynamical age ∼0.8 Myr. A comoving H I, molecular/H II shell is implied. The ring is inclined at 30° to the line of sight.

Whilst studying high velocities in the interstellar (IS) complex of M17/NGC 6618, Meaburn & Walsh (1981) considered several general mechanisms for the formation of high velocity phenomena and comoving H I and molecular/H II large-scale sheets. These included flows from ionization fronts, large-scale wind-driven cavities, internal supernova (SN) explosions, external SN explosions, localized wind-driven cavities, and radiation pressure. They note a variety of galactic H II regions have obvious SN remnants adjacent to them (Meaburn 1971) including the Rosette Nebula with the ∼40 pc diameter Monoceros remnant. A localized wind cavity in such a shell, as demonstrated in their fig. 8(a) is a possible model for the formation of the Rosette, although the question of the age discrepancy may still be difficult to resolve in this scenario.

A classic champagne flow (see Tenorio-Tagle 1979, and citations of this article) seems possible, if the Rosette had been formed in a thin cloud compared to its spatial extent on the sky. The actual Rosette cloud was found by Williams, Blitz & Stark (1995) using CO measurements to be an elongated structure of $1.65 \times 10^5$ M$_\odot$, with the Rosette Nebula at one end of a massive ridge (see Dent et al. 2009, for a more detailed map in CO 3–2). The close alignment of the ridge and Rosette cavity seem to be at odds then with a champagne flow, although projection effects may be important and Dent et al. (2009) note the motion of most clumps towards us, as opposed to the extent of the cloud on the sky. In any case, the thin cloud would also have to be very dense to focus the champagne flow away from the cloud and stop ablation over 2 Myr to such an extent that the cavity appears so relatively small. The origin of the extended ridge of the molecular cloud is not clear, although the shell of an ancient SN (as the Monoceros complex may be) or the remnant of an in-falling high velocity massive molecular cloud (Tenorio-Tagle et al. 1986) are possible explanations. Even so, high densities across the whole extent of such a large shell, not just in star-forming cores, would be hard to maintain over the time-scale required of tens of Myr.

In discussing M17, Clayton et al. (1985) suggest a breakout of the edge of an expanding bubble could be responsible for the physical structure and velocity dynamics in M17. A similar scenario may apply to the Rosette Nebula, with the stellar wind expanding into a rapidly declining density gradient, which could channel the champagne flow away from the rest of the shell, but significant difficulties arose in explaining the extensive velocity features.

A more recent solution involves a supersonically turbulent surrounding molecular cloud stalling the expansion of the wind-blown bubble (Geen et al. 2015). There, the emergence of an H II region from the host cloud is prevented by pressure forces, either thermal or turbulent, which also stop the cloud collapsing. In their model, based on that of Iffrig & Hennebelle (2015), the authors equate turbulent energy to gravitational energy for the low-mass cloud in question ($10^4$ M$_\odot$), whilst the thermal energy of the cold cloud is around 1 per cent of the gravitational energy. The kinetic energy is initially about 100 per cent of the gravitational energy, meaning that the cloud is globally supported by turbulence. The authors do not account for magnetic fields and it is not clear that the model cloud mass can be scaled to the 10 times larger Rosette Molecular Cloud, whilst also matching the velocity widths of ∼4 km s$^{-1}$ observed in the main part of the cloud (Williams, Blitz & Stark 1995).

Dent et al. (2009) also noted that the overall clumpy structure is very similar to the simulated images from smoothed particle hydrodynamics modelling of the effects of O star photoionization (see Dale, Clark & Bonnell 2007, and also fig. 2.). However, models such as presented by Geen et al. (2015) and Dale et al. (2007) do not crucially account for mass injection from the stellar wind, which we have shown elsewhere to be important (Rogers & Pittard 2013; Wareing et al. 2017a,b) and in the case of Dale et al. (2007) concern the external irradiation of a molecular cloud, whereas here we have internal irradiation. Tremblin et al. (2014) perform similar simulations to Geen et al. (2015) investigating the constraining effect of turbulent ram pressure (at Mach 1, 2, and 4), but do not consider the role of self-gravity or magnetic fields.

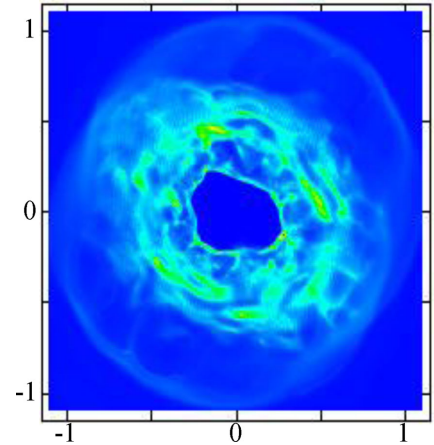

**Figure 2.** An example of simulated (optically thin) emission ($\rho^2 T$) from a 60 M$_\odot$ star in a thin sheet-like 17 000 M$_\odot$ molecular cloud, projected to approximately match the Rosette Nebula, produced from work presented elsewhere (Wareing, Pittard & Falle 2017a,b). The unit of distance is 50 pc. The central cavity in this example is approximately 25 pc in diameter.

To date, all of the explanations put forward for the apparent structure of the Rosette Nebula rely on further complexity over and above feedback in a molecular cloud (e.g. ejection events or highly supersonic turbulence). We instead wish to explore whether a simplified set of assumptions can reproduce a structure such as that found so uncomfortable by Bruhweiler et al. (2010). The idea of sheet-like configurations has been present in the literature for some considerable time, since at least the seminal work of Larson (1981), with, for example, Li & Smith (2005) recently proposing sheet-like structures based on the gravoturbulent model of molecular cloud formation. More recently, we have achieved molecular sheet-like cloud structures through magnetohydrodynamical simulations of thermally unstable diffuse clouds (Wareing et al. 2016). These structures have much in common with Larson (1981), for example apparent filamentary structure and trans-sonic velocities achieved as the thermal flows agglomerate into cooled molecular clumps. Our recent feedback simulations (Wareing et al. 2017a) have shown that such thin sheet-like structures can channel stellar winds away in highly asymmetric structures with very large ratios of major to minor axes, equal to and greater than that required by Bruhweiler et al. We show the possible emission from a structure formed by a 60 M$_\odot$ star in a low-mass molecular cloud in Fig. 2, in a simulation initial condition almost identical to that of the 40 M$_\odot$ star in Wareing et al. (2017a), but using the 60 M$_\odot$ star evolutionary track presented in Wareing et al. (2017b). That said, this recent magnetohydrodynamic feedback work and the matching non-magnetic case (Wareing et al. 2017b) have been limited to clouds with 17 000 M$_\odot$ of material. However, the Rosette Nebula resides in a much larger molecular cloud, in total around 165 000 M$_\odot$ (Williams, Blitz & Stark 1995). It is clear from our recent work that magnetic collimation of the mechanical stellar wind feedback (introduced as a source of mass and energy in the computation) can create elongated structures with surprisingly small cavities in a thin parent sheet (e.g.

Fig. 2), but does this model scale up to clouds like the Rosette complex and do the clouds remain relatively thin in the magnetic field direction compared to their other spatial extents? This manuscript considers these two questions, once questions over the importance of feedback from the massive star HD 46223 are resolved.

## 2 A NEW PROPER MOTION ANALYSIS

Previous work concluded that HD 46223 is not even associated with the Rosette Nebula (Zacharias et al. 2004). Highly accurate data from the *Gaia* satellite (Gaia Collaboration 2016b) has allowed a new exploration of this question, the answer to which could reveal a secondary estimate for the ages of the massive stars in NGC 2244 and also set an age limit for any models of the Nebula. Proper motions for five OB stars in NGC 2244 were obtained from *Gaia* Data Release 1 (Gaia Collaboration 2016a), including four stars with proper motions calculated from the combined *Hipparcos–Gaia* astrometry that lead to precisions of 0.05-0.09 milliseconds of arc per year (mas yr$^{-1}$) in right ascension and 0.04–0.06 mas yr$^{-1}$ in declination, and one star with proper motions calculated from combined *Tycho–Gaia* astrometry with precisions of 0.52 and 0.46 mas yr$^{-1}$ in right ascension and declination, respectively.

The proper motions show the five stars to be part of a coherent moving group. Subtracting the weighted mean proper motion of the cluster from the individual proper motions provides the motions of the stars in the cluster's reference frame. The stars HD 46223 (HIP 31149) and HD 46106 (HIP 31106) are moving away from the centre of the cluster and from each other with significantly larger velocities than the other stars in the cluster. Taking into account the uncertainties, the proper motions for both stars can be traced back to a single point in the cluster. The inferred velocities of HD 46223 and HD 46106 are, in the cluster reference frame, 0.19 ± 0.05 and 0.28 ± 0.07 mas yr$^{-1}$ in the directions indicated in Fig. 1(b). These equate to velocities of 1.38 and 2.05 km s$^{-1}$, assuming a distance of 1.53 kpc. Using a Monte Carlo simulation with $10^6$ iterations, we estimate that the two stars had coincident positions 1.73 (+0.34, −0.25) Myr (1σ uncertainties) in the past. We show the results of this proper motion analysis in Fig. 1(b) and data are available from https://doi.org/10.5518/311. At this time, they were seemingly collocated with the other members of NGC 2244, close to HD 46150 and the current centre of the cavity, as shown in Fig. 1(b). The implication is that HD 46223 was ejected from the cluster through stellar interactions at a time that matches the formation of the most massive stars and provides an age at which to investigate the results of our simulations, as discussed in the next section. SNe ejection would appear to be inconsistent given the inferred ages of the massive stars around 2 Myr, although not inconsistent if the massive O stars have ages of 6 Myr.

Bruhweiler et al. (2010) found no line broadening for IS lines of highly ionized species seen in the spectrum of HD 46223, as compared to significant and comparable levels of broadening from the other bright stars in NGC 2244 attributed to the effect of the wind-blown bubble. HD 46223 must have been ejected towards us on a vector that places it currently outside the wind-blown bubble and not contributing to its formation. Narrow IS features would be true, independent of whether HD 46223 was in front of or behind the expanding shell.

## 3 MODELS

Our recent work has highlighted the way that the thermal instability (Field 1965), under the influence of gravity and realistic magnetic fields (magnetic pressure), can drive the evolution of diffuse thermally unstable warm clouds from a pressure-supported quiescent low-density state to form high-density, cold (≤100 K) sheet-like structures that are filamentary in appearance perpendicular to the applied field (Wareing et al. 2016, hereafter Paper I). The winds and ionizing radiation from the massive stars that form through gravitational collapse of condensed molecular fragments in these sheets then strongly affect the environment into which core-collapse SNe subsequently explode. Often not accounted for, this pre-SNe feedback is particularly important to consider as it can introduce similar amounts of mass and energy to SNe, and can significantly alter the evolution of the SNe and their remnants (Wareing et al. 2017a,b, hereafter Papers II and III respectively).

In Paper II, realistic stellar wind feedback from a 40 M$_\odot$ star in a 17 000 M$_\odot$ corrugated sheet-like molecular cloud was able to evacuate a small central hole whilst the majority of stellar wind material was channelled away by the magnetic field perpendicular to the sheet-like structure. Sub-sonic, sub-Alvénic motions along the field lines in the initial diffuse cloud had led to the formation of the corrugated sheet, even in this case of initial thermal and magnetic pressure equivalence ($\beta_{\text{plasma}} = 1.0$). For more information, see Papers I and II. Further simulations of 32 and 60 M$_\odot$ stars in this sheet-like configuration in the 17 000 M$_\odot$ cloud, as shown in Fig. 2, have shown that tunnel-like structures with central cavities on the scale of the Rosette Nebula can be formed. In this paper, the question of whether this same effect can occur in a much larger molecular cloud, akin to the 165 000 M$_\odot$ cloud in which the Rosette Nebula resides, is now explored. This would provide a very elegant solution for the open questions surrounding the famous Rosette Nebula as discussed earlier.

### 3.1 Initial conditions

We present 3D, hydrodynamical and magnetohydrodynamical simulations of mechanical stellar feedback with self-gravity using the method recently presented elsewhere in Papers I, II, and III. A full description of the numerical approach using the MG code and how the physical models employed therein with differing values of the $\beta_{\text{plasma}}$ ratio of thermal to magnetic pressure lead to the formation of various molecular cloud structures is presented in Paper I. Mechanical feedback from a 15 M$_\odot$ star and from a 40 M$_\odot$ star into a 17 000 M$_\odot$ magnetically influenced sheet-like cloud formed through the action of thermal instability are shown in Paper II. Further, mechanical feedback from 15, 40, 60, and 120 M$_\odot$ stars into a 17 000 M$_\odot$ clumpy cloud formed by thermal instability without the influence of a magnetic field are shown in Paper III.

The Rosette Nebula is located in a much larger cloud, with a mass of approximately 165 000 M$_\odot$ (Williams, Blitz & Stark 1995). We have used the same technique as in Paper I, but with an initial domain twice as large (−3 to +3 in all directions) and an initial diffuse cloud with the same number density of atomic hydrogen throughout the medium of $n_H = 1.1$ cm$^{-3}$, but double the radius ($r = 100$ pc) such that the cloud now contains 135 000 M$_\odot$ of material. The cloud is seeded with random density variations – 10 per cent about this uniform initial density. Initial pressure is set according to the unstable equilibrium of heating and cooling at $P_{\text{eq}}/k = 4700 \pm 300$ K cm$^{-3}$ and results in an initial temperature $T_{\text{eq}} = 4300 \pm 700$ K. External to the cloud, the density is reduced by a factor of 10 to $n_H = 0.1$ cm$^{-3}$, but the external medium overpressured to match the cloud ($P_{\text{eq}}/k = 4700$ K cm$^{-3}$). Two simulations, one with no magnetic field ($\beta_{\text{plasma}} = \infty$) and one with initial magnetic-thermal pressure equality ($\beta_{\text{plasma}} = 1.0$) were then used to generate three new initial conditions, A, B, and C. As before, thermal instability causes the cloud to evolve into multiple clumps and contract in the hydrodynamic $\beta_{\mathrm{plasma}} = \infty$ case. In the magnetic field case with $\beta_{\mathrm{plasma}} = 1.0$, the cloud evolves into a thick disc perpendicular to the imposed field, with internal sheet-like structures that connect and form a filamentary network with diffuse voids. At late time, the cloud collapses into a thin corrugated sheet-like disc, much like the case of the low-mass cloud but formed over a longer time-scale with higher densities in the sheet.

### 3.2 Suite of simulations

Into these cloud structures, we now introduce feedback from a single massive star, approximating the dominance of HD 46150 in NGC 2244. We considering several cases to investigate the influence of the background magnetic field on the formation of the Rosette Nebula and the nature of the Rosette molecular cloud. The suite of simulations and range of parameter exploration carried out using the three initial conditions described above is detailed in Table 1. We introduce mechanical feedback through density and energy source terms in exactly the same manner as previously in Papers II and III. We choose when and where to inject the star by noting the time and location at which densities in each case rise to $100 \, \mathrm{cm}^{-3}$ – the density threshold often used for injection of stars in similar simulation work (e.g. Fogerty et al. 2016). We then wait a free-fall time (5.89 Myr) and switch on the star at $t_{\mathrm{inject}}$ at the qualifying location closest to the centre of the grid. In initial conditions A and B, the cloud has only had the higher densities associated with a molecular cloud for the previous 10 Myr – before that it was still a diffuse atomic cloud. The late magnetic case, initial condition C, is where a snapshot from the magnetic simulation has been selected when the cloud has collapsed through the thick disc stage into a 200-pc diameter thin sheet-like disc, resembling the initial magnetic condition used in Paper II. We show snapshots of these three initial conditions at the point of injecting feedback in Fig. 3, in order to clearly show the differences between the clouds before the onset of mechanical stellar feedback. We refer the interested reader to Paper I for a full description of the evolutionary processes that led to the formation of the clouds. The considerable difference in extent between the clouds should be noted. In the hydrodynamic case, the cloud is slowly collapsing under gravity and currently has a radius approximately half that of the initial diffuse cloud condition. In the magnetic case, the cloud is supported across the field lines against gravitational collapse and the cloud material begins to move along the field lines, forming initially a thick disc with thermal-instability driven internal structure and then after 46.4 Myr, this thick disc has collapsed into a thin sheet-like disc.

For the stellar evolution, 40 and 60 $\mathrm{M}_{\odot}$ non-rotating Geneva stellar evolution models calculated by Ekström et al. (2012) were used. These provide realistic, variable mass-loss rates and in combination with the work of Vink et al. (2000, 2001) allow the derivation of realistic wind velocity. These calculated mass-loss rates (mass source term) and wind velocities are shown in Fig. 4, along with energy injection rates (energy source term) and the total injected mass as a function of time from the birth (injection) of the star. The mass-loss rates of the stellar winds between 1.5 and 2.0 Myr remarkably closely bracket the derived rates of HD 46150 and HD 46223 presented by Howarth & Prinja (1989); $10^{-5.70}$ and $10^{-5.80} \, \mathrm{M}_{\odot} \, \mathrm{yr}^{-1}$, respectively. Factors of a few difference given the large uncertainties over the clumping factors in the winds of O and B stars are not unreasonable. Observationally derived terminal wind velocities are also similar: $3100 \, \mathrm{km \, s}^{-1}$ for HD 46223 and $3150 \, \mathrm{km \, s}^{-1}$ for HD 46150, compared to theoretical values of $\approx 3100 \, \mathrm{km \, s}^{-1}$, 1.5 Myr into the evolution of both 40 and 60 $\mathrm{M}_{\odot}$ stars. We compute each simulation for 2 Myr after the injection of the star. By this time, the 40 $\mathrm{M}_{\odot}$ star has injected a total of 1.3 $\mathrm{M}_{\odot}$ of material and $1.2 \times 10^{50}$ erg of energy into the cloud. The 60 $\mathrm{M}_{\odot}$ star has injected a total of 4.7 $\mathrm{M}_{\odot}$ of material and $4.4 \times 10^{50}$ erg of energy into the cloud. For the more massive star, this is about half an SN explosion's worth of mass and energy (taking a canonical explosion energy and ejecta mass of $10^{51}$ erg and 10 $\mathrm{M}_{\odot}$). We use star masses of 40 and 60 $\mathrm{M}_{\odot}$ to bracket the lower and upper mass estimates for HD 46150 and to avoid any inaccuracy that could be introduced whilst interpolating to an intermediate mass from the models presented by Ekström et al. (2012). These star masses also bracket empirical mass-loss rates and terminal wind velocities derived from observations (Howarth & Prinja 1989).

On injection of each star, enough mass is present in a small spherical region around the injection location such that the star mass can be 'removed' from the grid in order to make the star. The density and pressure at injection time zero are set to that of the average surrounding interclump cloud medium. This region is typically the same size, or slightly larger than, the source injection region (which has a radius $r = 0.0293$). It is very quickly dominated by the rapidly expanding wind, If instead the remaining mass is left in the injection region, the stellar wind rapidly and unrealistically cools and hence the feedback effects are underestimated. Simulation 4 represents the other extreme in plausible scenarios whereby the expelled star has the stronger wind. The 60 $\mathrm{M}_{\odot}$ star is given a velocity of $(v_x, v_y, v_z) = (2.19, 0.0, 0.0) \, \mathrm{km \, s}^{-1}$ such that the distance between the 40 and the 60 $\mathrm{M}_{\odot}$ stars at a stellar evolution time of 1.73 Myr corresponds to the observed separation of HD 46150 and HD 46223 at 1.53 kpc and an inclination angle to the line of sight of approximately $45°$, as suggested by the properties of the inclined ring derived from observations and discussed in the previous section. This simulation explores the importance, if any, of the role played by HD 46223 upon ejection from NGC 2244 approximately 1.73 Myr previously.

## 4 RESULTS

In this section, we present our results. We present both 2D slices through the computational volume and 3D contour and volume visualizations, created using the VISIT software (VisIt Collaboration 2012). The raw data can be obtained from https://doi.org/10.5518/311.

### 4.1 The spherical cloud simulation

In Fig. 5, we show the results of simulation 1, the hydrodynamic simulation of feedback from a 60 $\mathrm{M}_{\odot}$ star without magnetic field. Specifically, we show a density slice through the computational volume and a column density projection along the *y*-axis after 1.75 Myr of stellar evolution. It is clear to see that after 1.75 Myr of evolution, the stellar wind from the star at this initial location has not broken out of the cloud. However, considerable structure due to the action of the wind has been created within the cloud, including a hot bubble and a swept-up shell. Multiple dense protrusions exist within the bubble. These form from denser parts of the cloud which are over-run by the expanding shell, and they suffer hydrodynamic ablation due to the hot shocked stellar wind material that streams past. In this way, the bubble is being mass-loaded (cf. Rogers & Pittard 2013, and Paper III).

**Table 1.** Suite of simulations performed in this work.

| Number | IC | Cloud shape | Magnetic? | $t_{inject}$ (Myr) | Number of stars | Mass ($M_\odot$) | Location (code units) | Velocity (km s$^{-1}$) | Duration (Myr) |
|---|---|---|---|---|---|---|---|---|---|
| 1 | A | Spherical | N | 28.0 | 1 | 60 | (0.2, 0.0, −0.2) | (0.0, 0.0, 0.0) | 2.0 |
| 2 | B | Thick disc | Y | 36.8 | 1 | 60 | (0.2519, 0.0, −0.1055) | (0.0, 0.0, 0.0) | 2.0 |
| 3 | B | Thick disc | Y | 36.8 | 1 | 40 | (0.2519, 0.0, −0.1055) | (0.0, 0.0, 0.0) | 2.0 |
| 4 | B | Thick disc | Y | 36.8 | 2 | 40 | (0.2519, 0.0, −0.1055) | (0.0, 0.0, 0.0) | 2.0 |
|   |   |             |   |      |   | 60 | (0.2519, 0.0, −0.1055) | (2.19, 0.0, 0.0) |   |
| 5 | C | Thin disc | Y | 46.4 | 1 | 60 | (0.03, 0.0, −0.16) | (0.0, 0.0, 0.0) | 2.0 |
| 6 | C | Thin disc | Y | 46.4 | 1 | 40 | (0.03, 0.0, −0.16) | (0.0, 0.0, 0.0) | 2.0 |

Amazingly, the presence of the wind-blown bubble within the cloud is not revealed in images of the projected column density! In fact, there is barely any evidence of the effect of the stellar wind whatsoever in the column density projections. While dense gas is clearly pushed around by the action of wind, it seems to be displaced in a mostly radial direction from the star, and thus makes little impact on column density maps. In the figure, we colourmap the entire range of the projected column density data. It is true that a different range, possibly linear rather than log, and different colourmap may reveal other structure. We have investigated varying the logarithmic range using the same colourmap, but have been unable to produce an image which shows any clearer holes through the structure and prefer to show a range comparable to later figures. There are some faint low-density (blue) regions around the position of the central star, which could be accentuated by a lower upper range limit. The key point is that regardless of data presentation, the movement of material in the cloud has simply not been conducive to generating a Rosette-like hole through the structure along any line of sight, due to the radial nature of the wind expansion.

On all the other planes, the stellar wind has not reached the extent shown in Fig. 5. After a further 0.25 Myr, not shown here, the wind has still not broken out of the cloud. For these reasons, we conclude that the stellar wind from a 60 $M_\odot$ star located close to the centre of a high-mass unmagnetized cloud is not able to clear a central cavity, and thus fails to reproduce the observed structure of the Rosette Nebula. Of course one could place the star closer to the edge of the cloud, but this would make it even harder for the stellar wind to clear a cavity in projection. Stars with less cloud mass around them, including those closer to the edge of the cloud, would presumably behave more like the stars we have modelled in the case of a 17 000 $M_\odot$ unmagnetized cloud elsewhere (see Paper III). From that work, we find that a small evacuated cavity also does not form in a low-mass cloud. Again, the stellar wind expands in all directions, hindered by the other high-density clumps in the cloud, until eventually the whole side of the cloud is blown out. For these reasons, we conclude that the magnetic field has played an important role in the evolution of the Rosette Nebula. For a far more complete discussion of the hydrodynamic cloud–wind–SN interaction of stars with various masses, we refer the interested reader to Paper III.

### 4.2 Thick disc cloud simulations

In Fig. 6, we show the results of simulations 2, 3, and 4 modelling feedback from single and double stars in the thick disc molecular cloud, the evolution of which has been influenced by a magnetic field. Specifically, after 1.75 Myr of stellar evolution, we show density slices through the computational volume and projections in various directions during: simulation 2, feedback from a 60 $M_\odot$ star; simulation 3, feedback from a 40 $M_\odot$ star; and simulation 4, feedback from a stationary 40 $M_\odot$ star, and an ejected 60 $M_\odot$ star. Due to the different structure and nature of the parent molecular cloud, in all three simulations the stellar wind has been able to breakout of the cloud. Over the 1.5–2 Myr age range set by the proper motion study, the stellar wind bubbles expand further into the lower density regions surrounding the thick disc of the molecular cloud. The stellar winds have also cleared a large cavity in the molecular cloud, which is asymmetric in shape due to the initial distribution of dense material around the star and the effect of the magnetic field. Unlike in the unmagnetized case, this time the presence of the wind-blown bubble is clearly revealed in column density maps. When viewed in the plane of the thick molecular sheet, the breakout of the bubble is seen by the slight enhancement of the column density to the right of the sheet (see the second row in Fig. 6). When the thick sheet is in the plane of the sky, we instead see a ring of enhanced column density (which traces the swept-up shell), surrounding a central hole of reduced column density (which traces the low-density interior of the hot bubble). Clearly there are non-radial motions in this case.

The size of the central cavity in projection appears smaller as the breakout occurs through holes in the swept-up shell and not through a fully evacuated hole or tunnel (which is what happens in the case of a lower mass cloud – see Paper II). When viewed at an inclination to match the Rosette Nebula (fourth row), a ring of higher density is still visible, again surrounding a region of reduced column density. Higher density filamentary structures are also visible within this central region – it is not fully evacuated, as observed in the Rosette Nebula. The column density map can still be straightforwardly interpreted as the result of denser material existing around a central lower density region, with the filamentary structure in the column density map indicating the presence of dense filaments and protruding in front of, within, and behind the inner wind cavity. In the case of simulation 2, concerning feedback from the 60 $M_\odot$ star, the inner region with lower column density has a projected diameter of about 30 pc at 1.5 Myr, growing to about 36 pc at 1.75 Myr, and to about 45 pc at 2.0 Myr. The high column density ring has a projected diameter of about 55 pc at 1.5 Myr, growing to about 65 pc at 1.75 Myr, and to about 65-70 pc at 2.0 Myr. In the case of simulation 3, concerning feedback from the 40 $M_\odot$ star, the physical sizes are approximately 25 per cent smaller, but still considerably larger (a factor of 2) than the dimensions of the cavity and molecular ring in the Rosette Nebula. In the case of simulation 4, concerning double star feedback, the inner region with lower column density has a projected diameter of about 32 pc at 1.5 Myr, growing to about 38 pc at 1.75 Myr, and to about 45 pc at 2.0 Myr. The high column density ring has a projected diameter of about 60 pc at 1.5 Myr, growing to about 65 pc at 1.75 Myr, and to about 65–70 pc at 2.0 Myr. We find that our results are more sensitive to the choice of central star mass than to the addition of a secondary (albeit less powerful) wind source.

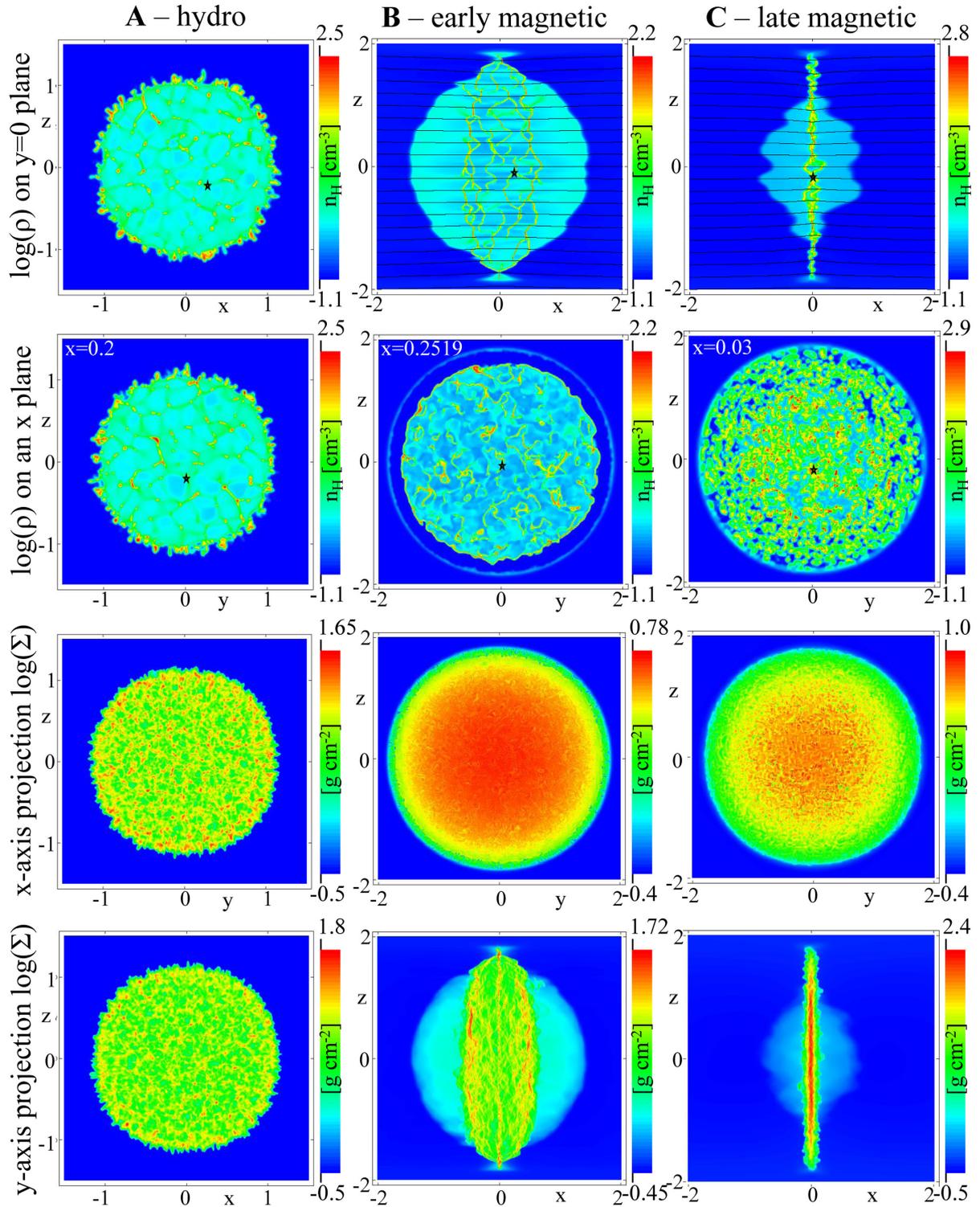

**Figure 3.** Initial conditions. The immediate pre-feedback appearance of the cloud in each of the three initial conditions A, B, and C. The black star symbol indicates the position of the star. Streamlines indicate direction of the magnetic field. The unit of distance is 50 pc. Refer to the text and Table 1 for more details. Raw data are available from https://doi.org/10.5518/311.

The tunnel-like cavity which formed in our previous work (Paper II) is not as clear when the cloud is thick and massive. It should be noted that simulations 2, 3, and 4 in a thick disc have only run for 2 Myr since the star formed. It remains quite possible that such a tunnel-like structure may also develop in these new simulations if they were advanced further in time. Nevertheless, they demonstrate that in a very massive thick-disc magnetized molecular cloud with internal interconnected filamentary structure, a central cavity can be cleared, but that this happens on spatial scales more than twice as large as that of the Rosette Nebula.

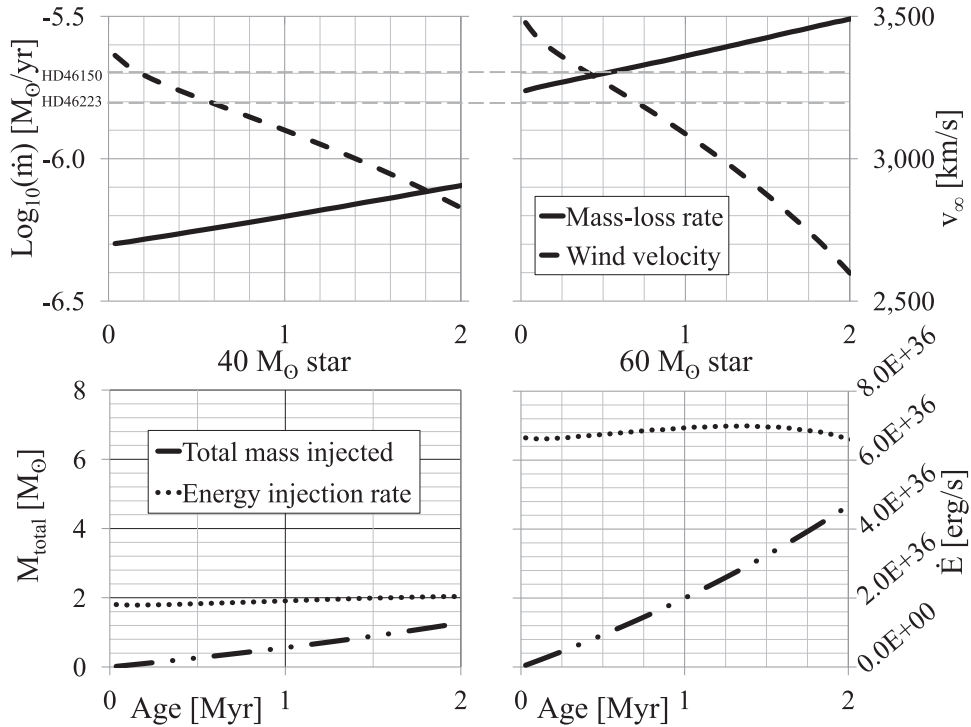

**Figure 4.** Stellar evolution tracks (Vink et al. 2000, 2001; Ekström et al. 2012) for 40 and 60 $M_\odot$ stars, showing mass-loss rate (solid line) and wind velocity (dashed line) on the upper graphs, and energy injection rate (dotted line) and total injected mass (dash double-dotted line) on the lower graphs. On the upper graphs, grey dashed horizontal lines represent observationally derived estimates of the mass-loss rates of HD 46150 and HD 46223 – see the text for more details. Raw data are available from https://doi.org/10.5518/311.

### 4.3 Sheet-like thin disc cloud simulations

In Fig. 7, we show the results of simulations 5 and 6 modelling feedback from single stars in a sheet-like thin disc molecular cloud, the formation of which has been influenced by a magnetic field. Specifically, the cloud is the same as that used in simulations 2, 3, and 4, but with injection of stars at a later time (10 Myr later) when the cloud has collapsed into a sheet-like thin disc. We show density slices through the computational volume and projections in various directions at: 1.75 Myr into the stellar evolution of simulation 5, with feedback from a 60 $M_\odot$ star; 1.75 Myr into the stellar evolution of simulation 6, with feedback from a 40 $M_\odot$ star; and 250 000 yr earlier, 1.5 Myr into the evolution of the same 40 $M_\odot$ star. Due to the thin disc structure and sheet-like nature of the parent molecular cloud, in both simulations the stellar wind has been able to breakout of the cloud, creating large bipolar wind structures that resemble those observed in Paper II from the same mass stars in an eight times less massive, but still sheet-like thin disc molecular cloud. Over the 1.5–2 Myr age range set by the proper motion study, the bipolar stellar wind bubbles expand considerably out into the lower density regions surrounding the molecular cloud. The stellar winds have each formed a small cavity in the molecular cloud, which is asymmetric in shape due to the initial distribution of dense material in the thin sheet-like disc around the location of the star and the effect of the magnetic field. The presence of the bipolar wind-blown bubble is only faintly revealed in column density maps viewed in the plane of the thin molecular sheet, indicated only by density enhancements in the compressed low-density cloud surrounding the molecular sheet (see the second row in Fig. 7). When the thin sheet is in the plane of the sky, we instead see indications of slightly enhanced column density (which traces the swept-up shell of the stellar wind ablating the molecular disc), surrounding a pronounced central cavity which traces the very low-density interior of the hot bubble. Clearly the motions in these simulations are predominantly channelled away from the molecular cloud by its thin, high-density nature.

The size of the central cavities in projection are much smaller than in the simulations with a thick disc molecular cloud. Clearly the fully evacuated central hole that has formed in these simulations is very much more like the case of a lower mass cloud presented in Paper II. Elongated tunnels are also forming in simulation 6 with the lower mass 40 $M_\odot$ star. The key factor in the formation of small evacuated cavities and elongated stellar wind tunnels is not the mass of the parent molecular cloud, but the thin nature of the cloud. This is a property which is influenced by the thermal-instability driven evolution in the presence of a magnetic field, although molecular clouds that are thin in one direction and extended perpendicular to this direction can be formed in other ways (e.g. cloud–cloud collisions and expanding SN remnants). When viewed at an inclination to match the Rosette Nebula (fourth row), a ring of higher density is still visible, again surrounding a fully evacuated cavity. In the case of simulation 5, concerning feedback from the 60 $M_\odot$ star, the cavity has a projected diameter of about 16 pc at 1.5 Myr, growing to about 20 pc at 1.75 Myr (shown in column 1 of Fig. 7), and to about 24 pc at 2.0 Myr. In the case of simulation 6, concerning feedback from the 40 $M_\odot$ star, the physical sizes are again approximately 25 per cent smaller, meaning that after 1.5 Myr, as shown in column 3 of Fig. 7, the asymmetric cavity has dimensions of 10 × 7.5 pc, closely matching that of the Rosette Nebula cavity. At later times, the cavity has grown larger than that observed, roughly by 20 per cent every 250 000 yr.

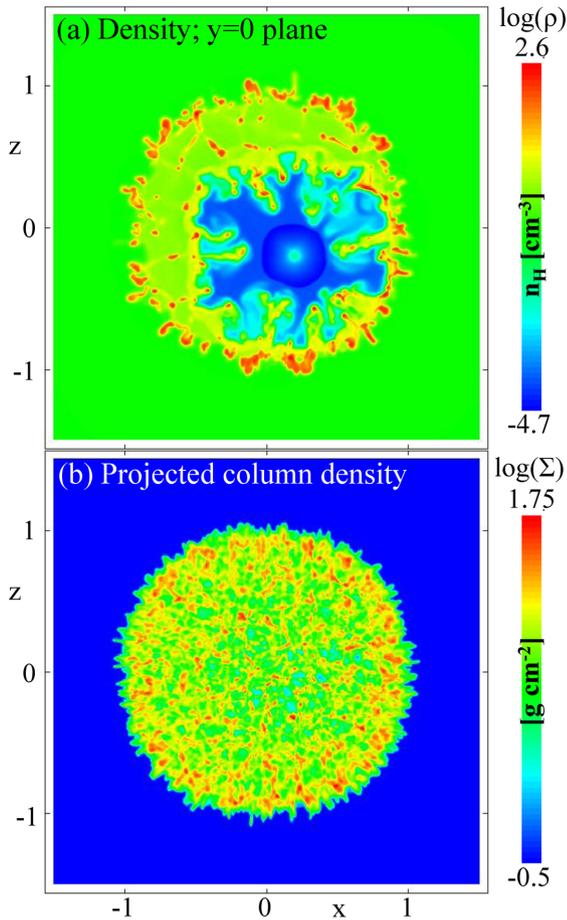

**Figure 5.** Cloud–wind interaction in simulation 1, the hydrodynamic case, 1.75 Myr into the evolution of the 60 M$_\odot$ star. Shown are the logarithm of mass density on the plane at $y = 0$ and column density projected along the *y*-axis of the computational volume. Length is scaled in units of 50 pc. Raw data are available from https://doi.org/10.5518/311.

Clearly the tunnel-like cavity which formed in our previous work (Paper II) has now formed in this work. The common denominator is that the molecular cloud is thin. Densities in the thin sheet-like disc of the cloud in this work are twice or more times higher than those in the cloud used in Paper II. This explains why the cavity is smaller in this more massive cloud. We have been able to show that in a very massive sheet-like magnetized molecular cloud, a central cavity on the same small scale as that observed in the Rosette Nebula can be cleared, surrounded by density enhancements which match the location of the peak of radio observations noted by Celnik (1985). In the next section, we go on to investigate the success of simulation 6 at reproducing this and other properties of the Rosette Nebula.

## 5 DISCUSSION

### 5.1 Comparison with observations

We first note that we have not run models over a large region of parameter space. Instead we have: (i) adopted very simple initial conditions for our models (that of a spherical diffuse atomic cloud of mass 135 000 M$_\odot$, an initial radius of 100 pc, and which was either magnetized, with $\beta_{plasma} = 1$, or unmagnetized); (ii) let the cloud contract under gravity while simultaneously forming structure through the thermal instability; and (iii) injected mechanical stellar wind feedback into the cloud. The feedback has come from either a single 60 M$_\odot$ star, a single 40 M$_\odot$ star, or from two stars (a centrally located 40 M$_\odot$ star, and an ejected 60 M$_\odot$ star). The stars form after a free-fall time once a specific density threshold is reached in the first two initial conditions and at a later time in the third initial condition once the magnetized cloud has collapsed into a thin sheet-like disc. We have not examined clouds of different initial mass, density, or shape, or other mechanisms for the formation of the cloud, nor other values of $\beta_{plasma}$, or other wind injection scenarios (from, e.g. different mass stars). It should therefore be clear that there remains a significant amount of parameter space for 'fine-tuning', even though we have been able to match observations of the Rosette Nebula in both qualitative and quantitative ways, as we will now explore.

The most obvious comparison to make between the observations of the Rosette Nebula and our models is that of the size of the central cavity. Celnik (1985) found a satisfactory fit to continuum radio maps at 4.75 GHz with a 'single shell hole model'. Scaled to the distance now favoured for HD 46150, the radius of the molecular shell (which is larger than that of the central cavity) is 6.7 pc. From the IPHAS observation in Fig. 1, we measure a radius of the central cavity between 4 and 5 pc at this distance, depending on the exact diameter chosen for the evacuated cavity. In comparison, the asymmetric cavity from simulation 6, 1.5 Myr into the evolution of the 40 M$_\odot$ star, is approximately 7.5 × 10 pc in total extent. The cavity is surrounded by a dense ring of swept-up wind and cloud material, with a radius of approximately 6–7 pc. Given this excellent agreement, we now focus our comparison with observations to simulation 6 at 1.5 Myr only.

We also obtain a good agreement for the gas velocity. We note that Bruhweiler et al. (2010) adopted an H II region expansion velocity of 13 km s$^{-1}$ from the observational data presented by Celnik (1985). In comparison with our simulations, at the inner edge of the thin sheet-like disc, where the wind is ablating the molecular material, we find velocities around 10–20 km s$^{-1}$. These reduce to a few km s$^{-1}$ in the thin sheet-like disc itself, in good agreement with the 21 cm radio observations of Kuchar & Bania (1993) that derived 4.5 km s$^{-1}$ in this region and the 4 km s$^{-1}$ of Williams, Blitz & Stark (1995).

Diffuse X-ray emission with a characteristic temperature around 9 × 10$^6$ K has also been detected from the cavity (Townsley et al. 2003). In our simulation, the gas immediately downstream of the stellar wind termination (reverse) shock(s) has a temperature in excess of 10$^8$ K. However, this gas is very low density. Further downstream the gas mixes with material ablated from the thin disc and subsequently reduces in temperature. At the contact discontinuity separating wind from cloud material the temperature of the shocked wind material reduces further, to below 10$^6$ K. Since the cooler parts of the shocked wind material are also the most dense we expect the X-ray emission to be dominated by such gas, which is consistent with the intermediate temperature measured by Townsley et al. (2003). A more detailed comparison could be made (e.g. Rogers & Pittard 2014), but this is beyond the scope of the current work.

Bruhweiler et al. (2010) note that emission from HD 46223, the massive star ejected from NGC 2244, is not line broadened, unlike the other massive stars in NGC 2244. The line broadening is caused by the expanding stellar wind bubble. The simplest interpretation to provide an alternative explanation for the size discrepancy and the missing wind problem without recourse to the extra complexity of stellar ejection events suggested by Bruhweiler et al. (2010) or highly supersonic turbulence employed in other works. We conjecture that the channelling away of the stellar wind from the cloud by the molecular sheet and the magnetic field leads to localized star formation to the south-east of the Nebula. Dense gas surrounding the Nebula in the plane of the sheet-like cloud may block emission from any triggered star formation in the north-west behind the Nebula. Alternatively, there may simply be a lack of material in this direction. We explore evidence for the orientation of the magnetic field in the next sub-section.

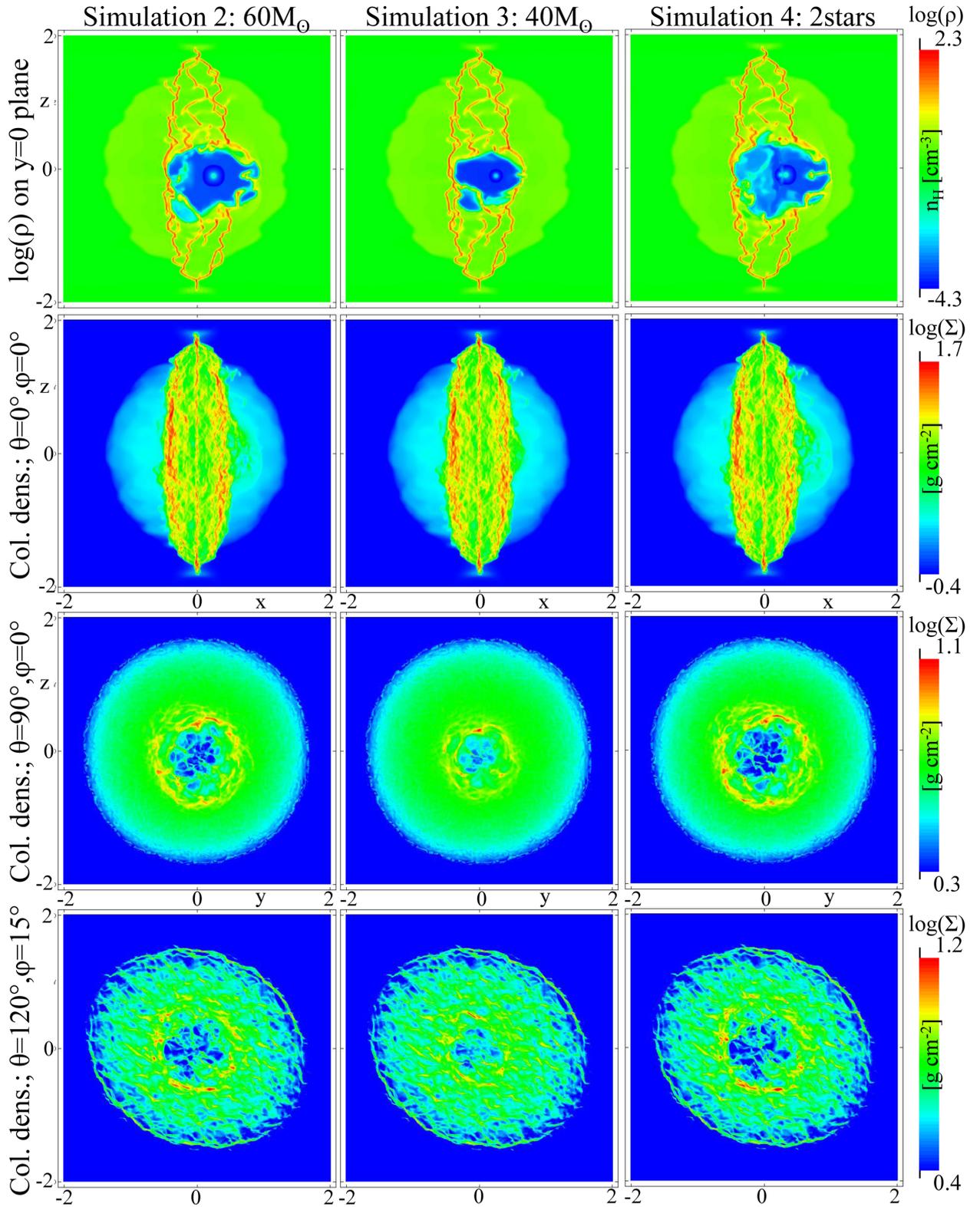

**Figure 6.** Cloud–wind interaction in the magnetic thick disc case. Shown are frames and projections from simulations 2, 3, and 4, all at 1.75 Myr into each simulation. Across the top row is the logarithm of mass density on the plane at $y = 0$, whilst on the lower rows are column density projected along the $y$-axis of the computational volume (second row), along the $x$-axis of the computational volume (third row), and inclined to match the Rosette Nebula (fourth row). Length is scaled in units of 50 pc. Raw data are available from https://doi.org/10.5518/311.

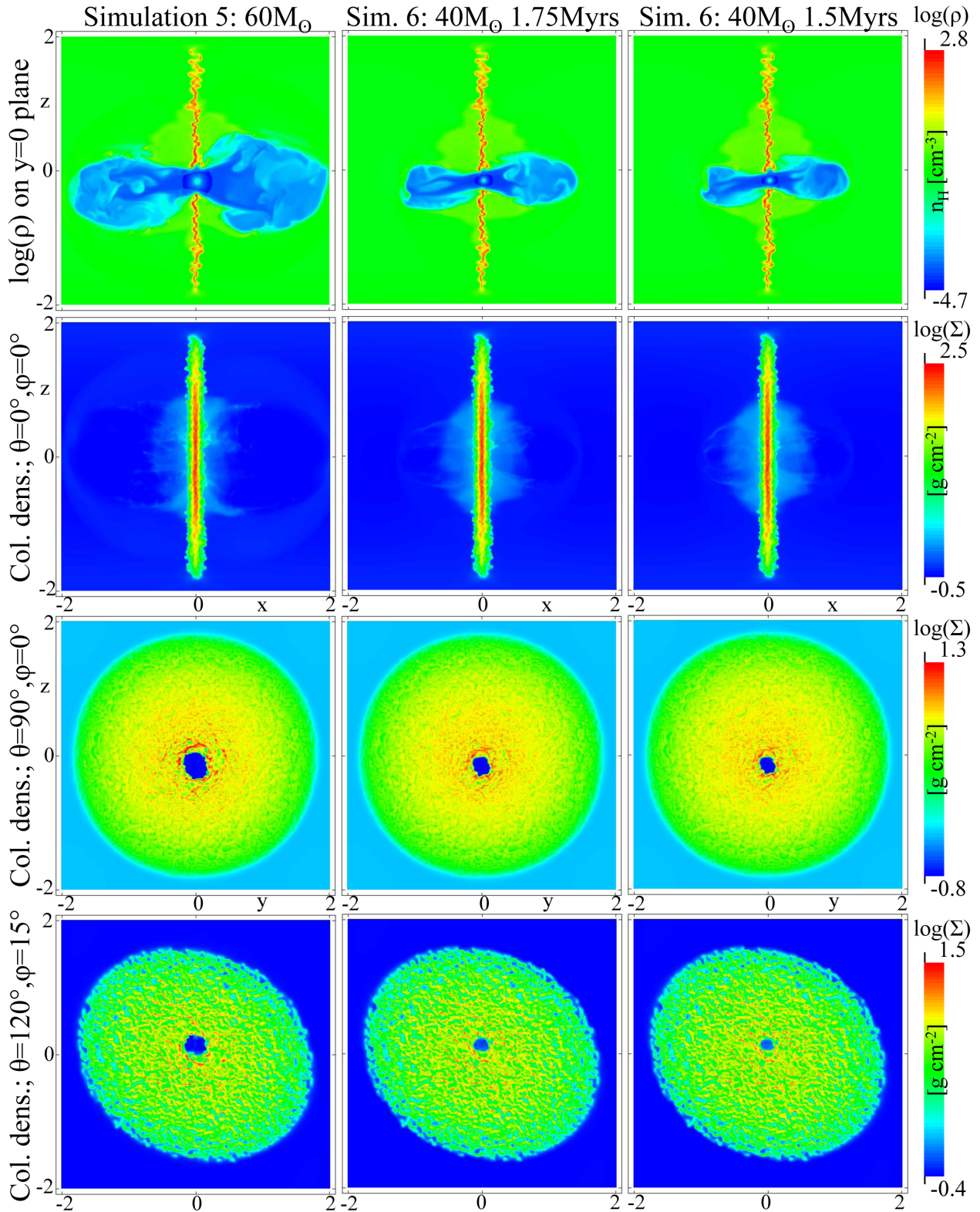

**Figure 7.** Cloud–wind interaction in the magnetic sheet-like thin disc case. Shown are frames and projections from simulations 5 and 6, at various times into each simulation. Across the top row is the logarithm of mass density on the plane at $y = 0$, whilst on the lower rows are column density projected along the $y$-axis of the computational volume (second row), along the $x$-axis of the computational volume (third row), and inclined to match the Rosette Nebula (fourth row). Length is scaled in units of 50 pc. Raw data are available from https://doi.org/10.5518/311.

### 5.2 Star formation and the magnetic field

The Rosette Nebula and its surroundings are a very complicated region of star formation. The Rosette Nebula itself is at one end of a ridge, which could explain the absence of any observed star formation to the north-east. *Herschel* column density maps reveal a filamentary structure to the Rosette Molecular Cloud, with all known infrared clusters lying at the intersection of these filaments (Schneider et al. 2012). These authors note that this is predicted by turbulence simulations, but we wish to emphasize that it is also consistent with our simulations. Our simulations generate the same characteristic line widths that are observed in the Rosette Molecular Cloud over a range of size scales, this being the very definition of Larson-like cloud turbulence (Larson 1981). In our case, however, this turbulence is a natural consequence of the gravitational collapse of a thermally unstable cloud, and is not imposed as an initial condition. Further, the structures formed in our scenario are stable and slow moving, generating high-density regions that persist for long enough for stars to form.

Simulations by Ibáñez-Mejía et al. (2016) of a magnetized, stratified, SN-driven interstellar medium (ISM), including diffuse heating and radiative cooling have also shown that self-gravity induces non-thermal motions as gravitationally bound clouds begin to collapse, approaching the observed relations between velocity dispersion, size, and surface density. These authors noted that in order to agree with observed star formation efficiencies, the process they model must be terminated by the early destruction of clouds, presumably from internal stellar feedback. We demonstrate the efficiency of precisely such destructive feedback effects in Papers II and III.

Schneider et al. (2012) emphasize that the star formation near the H II region does not show signs of being substantially different from that in the rest of the cloud, but that nevertheless locally induced star formation could be occurring at the interaction zone between the expanding H II region and the molecular cloud. The existence of a temperature gradient and a tentative age gradient of sources (Schneider et al. 2010) has previously been noted in the Rosette Molecular Cloud which is consistent with stellar wind triggering playing an important role. However, it is still not certain whether this is in fact the case, or whether the observed star formation solely results from the natural evolution of the molecular cloud.

In our thin sheet-like disc simulations, the channelling away of the stellar wind from the cloud due to the nature of the cloud and the magnetic field suggests that localized triggered star formation could occur to the south-east of the Nebula, along the direction of the imposed magnetic field in the simulation. This scenario is remarkably consistent with the *Planck* observation shown in Fig. 8, where a good agreement between the predicted field direction from the model and observation (Planck Collaboration 2016) is seen. Dense gas surrounding the Nebula in the plane of the sheet-like of this lack of line broadening is that even though it appears on the plane of the sky at the edge of the evacuated cavity, HD 46223 is in fact outside the stellar wind bubble. In our model with a narrowly collimated stellar wind bubble focused away from the parent molecular cloud, a wide range of cluster ejection vectors exist that would place HD 46223 outside and in front of the stellar wind bubble. In a more classical scenario of an expanding spherical wind bubble, it is considerably harder to explain this observation without recourse to a very high ejection velocity.

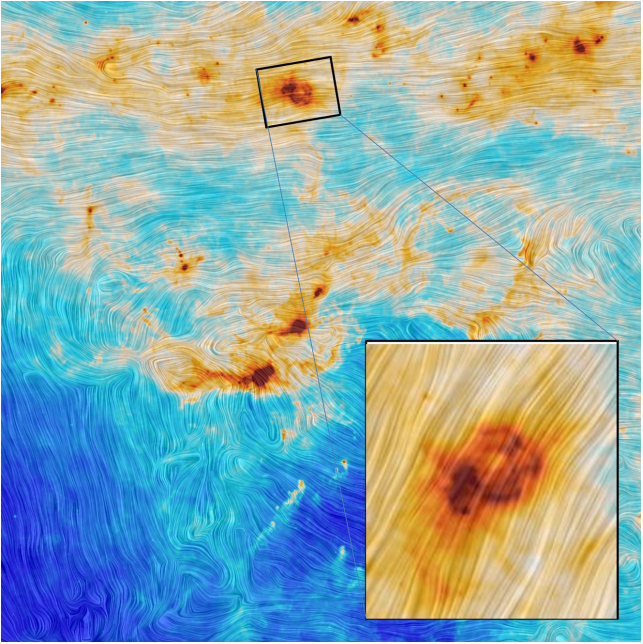

**Figure 8.** Image of magnetic field orientation, represented by the texture, with total intensity of dust emission shown in the colour scale, based on data from ESA's *Planck* satellite. The main image shows star formation and magnetic turbulence in the Orion Molecular Cloud. The expanded detail shows the Rosette Nebula and Rosette Molecular Cloud rotated to match Fig. 1. Copyright: ESA and Planck Collaboration. Reproduced with written permission from the European Space Agency.

It has been noted that densities across the Nebula are not supported by a simple shell model (Celnik 1985; Bruhweiler et al. 2010). Naively, column density through the Rosette Nebula should be constant across the Nebula (outside the cavity) if the cloud has a sheet-like morphology, or increasing towards the centre if the cloud has a spherical morphology. Observations (Kuchar & Bania 1993), assuming the emission is optically thin, show no clear evidence for increasing column density towards the centre of the Nebula, corresponding to a more sheet-like morphology and supporting the model presented here. In the magnetic simulations, compression of the sheet at the ablating inner edge of the cavity leads to weakly decreasing density with increasing radius, some evidence for which has been noted (Celnik 1985; Kuchar & Bania 1993).

In simulation 6, the stellar wind is able to create a central evacuated cavity that is far smaller than a typical Strömgren sphere, for the age of the star. Key to the formation of this structure is the presence of the magnetic field during the evolution of the parent molecular cloud. The thin sheet-like cloud and magnetic field have had chance to channel the stellar wind in this simulation. We had previously established that such a model could create tunnels and cavities in low-mass clouds (Paper II) and now we have naturally formed a central cavity of the same size as the Rosette Nebula after 1.5 Myr in a much larger cloud, confirming the model's ability cloud may block emission from any triggered star formation in the north-east behind the Nebula (or as noted this may be due to the Nebula being at one end of a ridge). To aid understanding, we show in Fig. 9 how the best-fitting snapshot at 1.5 Myr from simulation 6 can be rotated and inclined to align to the physical properties of the Rosette Nebula and the observed background magnetic field.

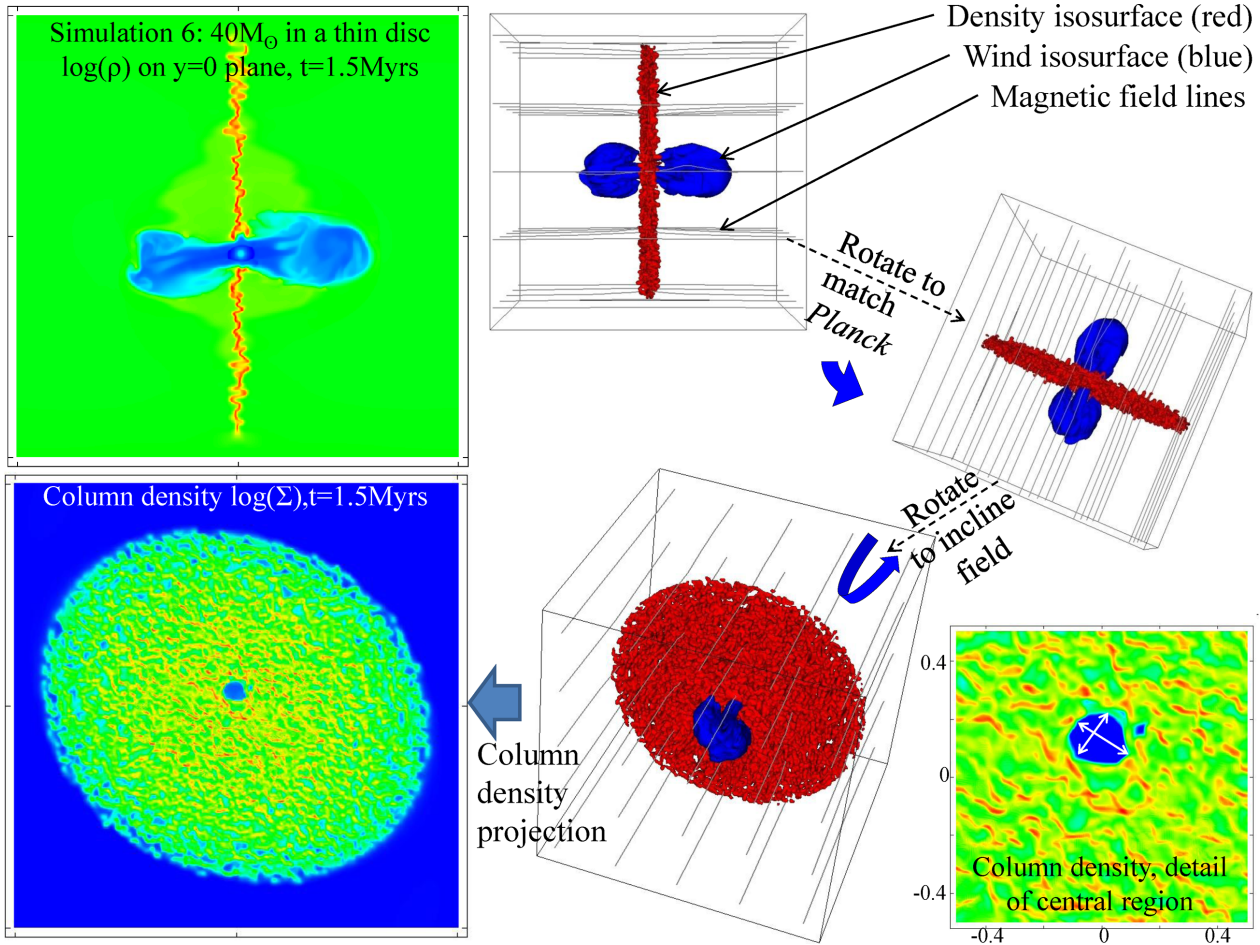

**Figure 9.** Rotation and inclination of the sheet-like thin disc molecular cloud simulation with feedback from a single 40 $M_\odot$ star, 1.5 Myr into the stellar evolution, in order to demonstrate the best match to the Rosette Nebula and to aid understanding. The detail of the central region shows the asymmetric cavity in the simulation, with dimensions 10 × 7.5 pc (length in the simulation is scaled in units of 50 pc).

### 5.3 Limitations

In this sub-section, we discuss limitations of this work. We have discussed other possible models that explain the physical properties of the Rosette Nebula at length in Introduction and refer the interested reader back to that section for comparison in light of our results.

First, we note that our simulated cloud resides in an isolated environment. Recent work (see Padoan et al. 2016, and the following series of papers) emphasizes the importance of field SN blast waves for cloud structure and dynamics. Gravitational and thermal instabilities within thin dense layers are thought to drive formation of filaments threaded by magnetic fields (e.g. Kudoh et al. 2007; Vázquez-Semadini et al. 2011; Van Loo, Keto & Zhang 2014). Recent work (Burge et al. 2016) investigating the effect of ambipolar diffusion and decaying turbulence on infinitely long, isothermal, magnetically sub-critical filaments in two dimensions found that, by perturbing the equilibria with decaying velocity perturbations, these equilibrium filaments are dynamically stable. Such structures are not inconsistent with those from large-scale gravitational instability forming sheets intersecting in filaments. We have no doubt that gravitational instability will take over the evolution of these simulations as the density increases, but thermal instability appears to provide an equally appealing model for the initial evolution of molecular clouds. Further work is required on all aspects of this.

Secondly, we note that our current models do not include radiative feedback. Such feedback can drive gas from clusters (e.g. Dale, Ercolano & Bonnell 2012; Walch et al. 2012; Geen et al. 2015; Raskutti, Ostriker & Skinner 2016; Shima, Tasker & Habe 2017; Howard, Pudritz & Harris 2017), and needs to be included in future models in order to obtain a clearer picture of the feedback effects of massive stars.

### 6 SUMMARY AND CONCLUSIONS

We have explored and simulated an evolutionary scenario for the Rosette Nebula and its parent molecular cloud that reproduces qualitatively and quantitatively the structure of the Rosette. We have found a solution for the long-standing age discrepancy regarding the central cavity (see Bruhweiler et al. 2010, and references therein) that has hampered the understanding of the Rosette. We have achieved this by combining the natural thermal instability of the diffuse ISM with a background magnetic field and gravity in order to form a molecular cloud with a thin sheet-like structure, rather than something which is more spherical in nature. The feedback mechanism here differs considerably from the traditional picture, where wind material is confined within the parent molecular cloud by high densities all around the star. We also demonstrate that a purely hydrodynamic mechanism and a thick sheet-like structure are unable to recreate the structure of the Rosette Nebula, and hence discover that magnetic fields leading to the formation of thin sheets play a key role.

In the magnetic case, the field affects the formation of the molecular cloud and allows the subsequent stellar wind(s) to breakout of the thin sheet-like cloud and create a small central cavity, even in the high-mass cloud case considered here ($135\,000\,M_\odot$). We have previously demonstrated that this mechanism also works in a low-mass cloud case ($17\,000\,M_\odot$: Wareing et al. 2017a). In projection, the structure closely resembles the Rosette Nebula, with the required magnetic field matching the observed alignment and inclination. Our model also largely resolves the discrepancy between the dynamical age of the Nebula and the stellar ages. A new proper motion study of the most massive stars in NGC 2244 has revealed the ejection of HD 46223 from the centre of the cluster and set an age of 1.73 ($+0.34$, $-0.25$) Myr ($1\sigma$ uncertainty) for the Rosette Nebula. Our model reproduces the observed size of the cavity in the Rosette Nebula, 1.5 Myr into the evolution of feedback from a $40\,M_\odot$ star into a thin sheet-like disc. Our model can also explain the position and localized nature of any stellar wind affected star formation in the Rosette Molecular Cloud as being triggered by the bipolar structure formed by the stellar wind narrowly focused by the high densities in the thin sheet-like disc.

The model presented here starts from a diffuse, thermally unstable atomic cloud with realistic heating and cooling prescriptions in the presence of a magnetic field. The model does not require the extra complexity of individual stellar ejection events or a turbulent driving mechanism for the formation of the molecular cloud.

Further work is now required to elucidate the important effects of radiative feedback (which we are yet to include) and further explore the role of the magnetic field in the formation of molecular clouds. The observed filamentary appearance of molecular clouds, combined with the ubiquity of magnetic fields, the formation of clouds by collisions, the propensity of turbulence to form interconnecting sheets, and the mechanism highlighted here reinforces the fact that sheet-like molecular cloud structures must be common and can be formed in different ways, even if gravitational instability takes over in all cases and leads to the formation of very high-density clumps and eventually stars. We have highlighted here that feedback into a sheet-like morphology is very different from that into a spherical cloud and that it can resolve long-standing questions over the structure and age of the Rosette Nebula. Moreover, this issue may not be unique to the Rosette, but may also apply to other IS bubbles. We plan to explore such issues in future works.

## ACKNOWLEDGEMENTS

This work was supported by the Science and Technology Facilities Council (STFC, Research Grants ST/L000628/1 and ST/P00041X/1). The calculations for this paper were performed on the DiRAC 1 Facility at Leeds jointly funded by STFC, the Large Facilities Capital Fund of BIS and the University of Leeds and on other HPC facilities at the University of Leeds. These facilities are hosted and enabled through the ARC HPC resources and support team at the University of Leeds (A. Real, M. Dixon, M. Wallis, M. Callaghan, and J. Leng), to whom we extend our grateful thanks.

The DiRAC Data Centric system at Durham University was also used, operated by the Institute for Computational Cosmology on behalf of the STFC DiRAC HPC Facility (www.dirac.ac.uk). This equipment was funded by a BIS National E-infrastructure capital grant ST/K00042X/1, STFC capital grant ST/K00087X/1, DiRAC Operations grant ST/K003267/1, and Durham University. DiRAC is part of the National E-Infrastructure. This work has made use of data from the European Space Agency (ESA) mission *Gaia* (http://www.cosmos.esa.int/gaia), processed by the *Gaia* Data Processing and Analysis Consortium (DPAC, http://www.cosmos.esa.int/web/gaia/dpac/consortium). Funding for the DPAC has been provided by national institutions, in particular the institutions participating in the *Gaia* Multilateral Agreement. We acknowledge an extensive and valuable review from a Science Advances anonymous referee which considerably increased the scope of the original work. We also acknowledge useful comments from colleagues in the School of Physics and Astronomy in Leeds (T. W. Hartquist, S. Van Loo, M. Hoare, R. Oudmaijer, and K. Johnston), collaborators (A. A. Zijlstra), and audience members at presentations elsewhere. Finally, we gratefully acknowledge the positive review and warm comments from the MNRAS Reviewer, Dr Frederick C Bruhweiler. Data for the figures in this paper are available from http://doi.org/10.5518/311. VISIT is supported by the Department of Energy with funding from the Advanced Simulation and Computing Program and the Scientific Discovery through Advanced Computing Program.

## REFERENCES

Bonatto C., Bica E., 2009, MNRAS, 394, 2127
Bruhweiler F. C., Freire Ferrero R., Bourdin M. O., Gull T. R., 2010, ApJ, 719, 1872
Burge C., Van Loo S., Falle S. A. E. G., Hartquist T. W., 2016, A&A, 596, A28
Celnik W. E., 1985, A&A, 144, 171
Clayton C. A., Ivchenko V. N., Meaburn J., Walsh J. R., 1985, MNRAS, 216, 761
Dale J. E., Clark P. C., Bonnell I. A., 2007, MNRAS, 377, 535
Dale J. E., Ercolano B., Bonnell I. A., 2012, MNRAS, 424, 377
Dent W. R. F. et al., 2009, MNRAS, 395, 1805
Drew J. E. et al., 2005, MNRAS, 362, 753
Ekström S. et al., 2012, A&A, 537, A146
Field G. B., 1965, ApJ, 142, 531
Fogarty E., Frank A., Heitsch F., Carroll-Nellenback J., Haig C., Adams M., 2016, MNRAS, 460, 2110
Freyer T., Hensler G., Yorke H. W., 2003, ApJ, 594, 885
Freyer T., Hensler G., Yorke H. W., 2006, ApJ, 638, 262
Gaia Collaboration et al., 2016a, A&A, 595, A2
Gaia Collaboration et al., 2016b, A&A, 595, A1
Geen S., Hennebelle P., Tremblin P., Rosdahl J., 2015, MNRAS, 454, 4484
Howard C. S., Pudritz R. E., Harris W. E., 2017, MNRAS, 470, 3346
Howarth I. D., Prinja R. K., 1989, ApJS, 69, 527
Ibáñez-Mejía J. C., Mac Low M.-M., Klessen R. S., Baczynski C., 2016, ApJ, 824, 41
Iffrig O., Hennebelle P., 2015, A&A, 576, A95
Kuchar T. A., Bania T. M., 1993, ApJ, 414, 664
Kudoh T., Basu S., Ogata Y., Yabe T., 2007, MNRAS, 380, 499
Larson R. B., 1981, MNRAS, 194, 809
Li J. Z., Smith M., 2005, AJ, 130, 2757
Mahy L., Nazé Y., Rauw G., Gosset E., De Becker M., Sana M., Eenens P., 2009, A&A, 502, 937
Martins F., Mahy L., Hillier D. J., Rauw G., 2012, A&A, 538, A39
Meaburn J., 1971, Ap&SS, 13, 110
Meaburn J., Walsh J. R., 1981, Ap&SS, 74, 169
Padoan P., Pan L., Haugbølle T., Nordlund Å., 2016, ApJ, 822, 11

Planck Collaboration XXXIV, 2016, A&A, 586, A137
Raskutti S., Ostriker E. C., Skinner M. A., 2016, ApJ, 829, 130
Rogers H., Pittard J. M., 2013, MNRAS, 431, 1337
Rogers H., Pittard J. M., 2014, MNRAS, 441, 964
Schneider N. et al., 2010, A&A, 518, L83
Schneider N. et al., 2012, A&A, 540, L11
Schneps M. H., Ho P. T. P., Barratt A. H., 1980, ApJ, 240, 84
Shima K., Tasker E. J., Habe A., 2017, MNRAS, 467, 512
Tenorio-Tagle G., 1979, A&A, 71, 59
Tenorio-Tagle G., Bodenheimer P., Rozyczka M., Franco J., 1986, A&A, 170, 107
Townsley L. K., Feigelson E. D., Montmerle T., Broos P. S., Chu Y.-H., Garmire G. P., 2003, ApJ, 593, 874
Tremblin P. et al., 2014, A&A, 568, A4
Van Loo S., Keto E., Zhang Q., 2014, ApJ, 789, 37
Vázquez-Semadini E., Banerjee R., Gómez G. C., Hennebelle P., Duffin D., Klessen R. S., 2011, MNRAS, 414, 2511
Vink J., de Koter A., Lamers H. J. G. L. M., 2000, A&A, 362, 295
Vink J., de Koter A., Lamers H. J. G. L. M., 2001, A&A, 369, 574
VisIt Collaboration, 2012, High Performance Visualisation – Enabling Extreme-Scale Scientific Insight, CRC Press, Boca Raton, FL, p. 357
Walch S. K., Whitworth A. P., Bisbas T., Wünsch R., Hubber D., 2012, MNRAS, 427, 625
Wang J., Townsley L. K., Feigelson E. D., Broos P. S., Getman K. V., Román-Zúniga C. G., Lada E. A., 2008, ApJ, 675, 464
Wareing C. J., Pittard J. M., Falle S. A. E. G., Van Loo S., 2016, MNRAS, 459, 1803 (Paper I)
Wareing C. J., Pittard J. M., Falle S. A. E. G., 2017a, MNRAS, 465, 2757 (Paper II)
Wareing C. J., Pittard J. M., Falle S. A. E. G., 2017b, MNRAS, 470, 2283 (Paper III)
Weaver R., McCray R., Castor J., Shapiro P., Moore R., 1977, ApJ, 218, 377
Williams J. P., Blitz L., Stark A. A., 1995, ApJ, 451, 252
Wright N. J., Drew J. E., Mohr-Smith M., 2015, MNRAS, 449, 741
Zacharias N., Monet D. G., Levine S. E., Urban S. E., Gaume R., Wycoff G. L., 2004, BAAS, 36, 1418

This paper has been typeset from a TeX/LaTeX file prepared by the author.